\documentclass[12pt]{article}
\usepackage[T1]{fontenc}
\usepackage[utf8]{inputenc}
\usepackage{lmodern}
\usepackage{amsmath}
\usepackage{amssymb}
\usepackage{graphicx}
\usepackage{caption}
\usepackage{subcaption}
\usepackage{auto-pst-pdf}
\usepackage{multirow}
\usepackage{float}

\title{Particle Mesh Ewald's Method and Non-Interacting Dyon Gas}
\author{Motahareh Kiamari$^{1}$, Sedigheh Deldar$^{1}$ \\ \\
$^1$Department of Physics, University of Tehran,\\
\small P.O. Box 14395/547, Tehran 1439955961,
Iran.}
\date{}
\begin{document}
\maketitle

\begin{abstract}
We study the free energy of a quark-antiquark pair near the deconfinement temperature by particle mesh Ewald's method for non-interacting dyon ensemble. 
We show that the free energy of the quark-antiquark pair increases linearly by increasing the distance between them.
The string tension decreases by increasing the temperature, as expected.  
 
\end{abstract}

\section{Introduction}
\label{sec:intro}

Finding a mechanism to describe  quark confinement has been one of the interesting subjects in particle physics since the development of QCD for describing the strong interaction. Studying and understanding the structure of QCD vacuum is the main goal. Many papers have suggested magnetic monopoles, vortices, instantons, dyons or KvBLL calorons as the main candidates for QCD vacuum constituents and have tried to explain quark confinement by these objects (see \cite{12,13} as examples). 

Diakonov and Petrov \cite{1}, studied quark confinement by non-interacting ensemble of KvBLL calorons and their constituents dyons. They found the critical temperature of confinement-deconfinement transition phase by these structures. After that, these structures attracted a lot of interests and some people tried to add some kind of interactions between dyons to improve this model. However, Bruckmann and \textit{et al}. \cite{2} showed that the metric Diakonov and Petrov introduced for their model, is only positive definite for dyons of different charges or for dyons of the same charge at separation larger than $ \frac{2}{\pi T} $ in SU(2) gauge group. They \cite{3} presented a numerical method named after P. P. Ewald \cite{4} and used dyons as the structures of the QCD vacuum. Although Ewald's method was originally developed for Coulombic interactions, it was soon extended to any other long-range interactions. The key idea of this method is to split the Coulombic term into an exponentially “short-range part” and a smooth “long-range part”. The former term converges but the latter diverges, therefore one should calculate it in Fourier space. This method can control the finite volume effects, efficiently.

Although a dyon is an SU(2) object but it can be considered as a U(1) object when observed from large distances. Therefore, it carries magnetic and electric charges and  Coulombic  magnetic and electric fields at large distances. Hence, Bruckmann \textit{et al}. \cite{3} applied Ewald's method to non-interacting ensemble of dyons. Particle mesh Ewald's method \cite{9} is an alternative numerical method for this type of calculations. It is more efficient and less time consuming than the simple Ewald's method. Assigning the griding charges to a lattice and computing the “long-range part” is the main difference between this method and the original Ewald's method which will be explained in detail in section 4. The CPU time of the performance of the particle mesh Ewald's method is of the order of $N\log N$ while the CPU time of the simple Ewald is of the order of $N^{\frac{3}{2}}$, where N is the dyon number.

Our aim in this article is to find the potential between a pair of qaurk antiquark by Particle Mesh Ewald's method in a non-interacting Dyon gas. What we have achieved is a linear potential and therefore confinement and also decreasing string tension by increasing the temperature with a method much less expensive compared with the original Ewald's method. In addition to this physical results we get, testing Particle Mesh Ewald's method seems to be a valuable task since it may be used for interacting dyon gas where Ewald's method is not efficiently applicable because of the expensive computer running time. It had been used by chemists for Chemistry problems but not by physicists, as far as we know. 
In the Particle Mesh Ewald's method, we use the charges assigned to the mesh points and then we apply the simple Ewald's method to the new charges. We fix the number of mesh points for all simulations and do our calculations in approximately fixed temperature near the deconfinement phase. Our results represent linear rising of free energy of a quark-antiquark pair by increasing their distances.  

The paper is organized as the following. In Section \ref{sec:dyon}, some features of dyons are introduced and the Polyakov loop correlator is derived. Section \ref{sec:em} introduces Ewald's method, briefly. In Section \ref{sec:pme}, particle mesh Ewald method is described. And in Section \ref{sec:results}, we introduce the setup of our simulations and the numerical results are presented. 

\section{Dyon ensemble for SU(2) Yang-Mills theory}
\label{sec:dyon}

KvBLL caloron, found by Kraan and van Baal \cite{5}, as well as Lee and Lu \cite{6}, is the periodic instanton solution of finite temperature Yang Mills theory. This solution consists of dyons which are the magnetic monopoles as well as electric charge. Dyons are originally SU(2) solutions of Yang-Mills theory and generally non-Abelian objects. But they are observed as Abelian objects in the far-field limit, where the distances to the center of dyons are large. In fact, the temporal gauge field and the magnetic and electric fields are Abelian along the third direction, color direction in SU(2),

\begin{equation}
A_{4}\rightarrow 2\pi\omega T  \sigma_{3},
\label{A4}
\end{equation}

\begin{equation}
\pm B=E\rightarrow \frac{q}{r^{2}} \sigma_{3}, 
\label{ebfield}
\end{equation}
where T is the temperature and $ \sigma_{3}=diag(+1,-1) $ is the third Pauli matrix and the Cartan generator of SU(2). The parameter $ \omega $ is the holonomy, and is related to the asymptotic Polyakov loop,

\begin{equation}
P(\textbf{r})=\frac{1}{2}Tr\left(\exp\left(i\int_{0}^{1/T} dx_{4}A_{4}\left(x_{4},\textbf{r}\right) \right) \right) \rightarrow \frac{1}{2}Tr\left(\exp\left(2\pi i\omega\sigma _{3}\right)\right)=\cos\left(2\pi\omega\right).  
\label{polyakovloop}    
\end{equation}  
\\The holonomy is the order parameter of the confinement-deconfinement transition. Maximally nontrivial holonomy, $ \omega=\frac{1}{4} $, specifies the confined phase, $ P(\textbf{r})\rightarrow 0 $. The trivial holonomy specifies the deconfined phase, $ P(\textbf{r})\rightarrow \pm 1 $.
To obtain the magnetic and the electric fields of equation (\ref{ebfield}), the Abelian gauge fields of dyon are defined as

\begin{align}
a_{4}\left(\textbf{r};q\right)=\frac{q}{r},  a_{1}\left(\textbf{r};q\right)=-\frac{qy}{r\left(r-z\right)},  a_{2}\left(\textbf{r};q\right)=+\frac{qx}{r\left(r-z\right)},  a_{3}\left(\textbf{r};q\right)=0.
\label{dyonpotential}
\end{align}
\\The plus and minus signs in equation (\ref{ebfield}), specify the self-dual and anti-self-dual equations, respectively. There are two dyons in SU($2$) gauge group, named M and L, with the possible magnetic charge $ q_{m}=\pm 1 $. As a consequence of self-duality, the electric charges have the same signs . There are also two anti-dyons, named $ \bar{M} $ and $ \bar{L} $, with the magnetic and the electric charges of unit value and the different signs as the consequence of anti-self-duality. Since we have studied only dyons, and the magnetic and electric charges of dyon are equal, one can consider dyons as the particles with the Abelian electric charge of $ \pm 1 $. It should be noticed that the L dyons can be obtained by replacing $ 2\omega $ with $ 1-2\omega $ in equation (\ref{A4}), but because we study the confined phase with $ \omega=\frac{1}{4} $, two dyons have the same topological charges and actions. 

The Polyakov loop correlator yields the free energy of a static quark-antiquark pair as a function of their separation $d$, 
\begin{align}
F_{\bar{Q}Q}(d)=-T\ln \left\langle P(\textbf{r})P^{\dag} (\textbf{r}')\right\rangle , d\equiv \lvert \textbf{r}-\textbf{r}' \rvert ,
\end{align}
and the expectation values of observables \textit{O} using path integrals

\begin{equation}
\langle O \rangle=\frac{1}{Z} \int \left( \prod_{k=1}^{n_{D}} d^{3}r_{k} \right) O\left( \left\lbrace \textbf{r}_{k}\right\rbrace \right) \exp \left[S\left( \left\lbrace \textbf{r}_{k}\right\rbrace \right)\right]  
\label{evofO}
\end{equation}
where \textit{Z} is the partition function 
\begin{equation}
Z=\int \left( \prod_{k=1}^{n_{D}} d^{3}r_{k} \right) \exp \left[S\left( \left\lbrace \textbf{r}_{k}\right\rbrace \right)\right].  
\label{partitionfun}
\end{equation}
Hence, one should obtain the Polyakov loop and the effective action of the dyon ensemble. Keeping in mind that the original objects we study are calorons which  are neutral objects consisting of  two dyons with opposite charges in SU(2) gauge group, we should construct an ensemble of equal number of M and L dyons to have a neutral ensemble. Using $a_4$ of equation (\ref{dyonpotential}) in the Polyakov loop of equation (\ref{polyakovloop}), for 2K dyons  

\begin{align}
P(\textbf{r})=\cos\left(2\pi\omega+\frac{1}{2T}\Phi(\textbf{r})\right),  P(\textbf{r})| _{\omega=1/4 } =-\sin\left(\frac{1}{2T}\Phi(\textbf{r})\right),  
\end{align}
\begin{equation}
\Phi(\textbf{r})\equiv \sum_{i=1}^{2K}\frac{q_{i}}{\lvert \textbf{r}-\textbf{r}_{i}\rvert }.  
\label{phi}
\end{equation}
\\For non-interacting ensemble, both M and L dyons have the same constant actions, which can be factored out in equation (\ref{evofO}) and then the free energy can be obtained.

\section{Ewald's Method}
\label{sec:em}

The first step in doing Ewald's method is to mimic the space by a basic cell, the “super cell”, and then to copy it in all directions. The physical system is restricted to be located in the super cell and a finite number of $ n_{D} $ dyons are placed in it randomly. The copies of super cell contain the copies of $ n_{D} $ dyons and they help to apply the periodic boundary condition to decrease the finite size volume effect. The second step is to split the $ \frac{1}{r^{p}} $ term into an exponentially “short-range part” and a smooth “long-range part”, where $ p\in \mathbb{R}$, $p\geq 1 $. With the Euler gamma function 
\begin{equation}
\Gamma(z)=\int_{0}^{\infty } t^{z-1}\exp (-t) dt=r^{2z} \int_{0}^{\infty } t^{z-1}\exp (-r^{2}t) dt
\end{equation} 
and the Fourier integral expansion of the three-dimensional Gaussian distribution
\begin{equation}
\exp\left(-\textbf{r}^{2}t\right)=\left(\frac{\pi}{t}\right)^{3/2}\int_{0}^{\infty}d^{3}\textbf{u}\exp\left(-\pi^{2}\textbf{u}^{2}/t\right)\exp\left(-2i\pi\textbf{u}.\textbf{r}\right).
\end{equation} 
With some mathematical operations, the $ \frac{1}{r^{p}} $ term is
\begin{equation}
\frac{1}{r^{p}}=\frac{\pi ^{3/2}}{\left(\sqrt{2}\lambda\right)^{p-3}} \int d^{3}\textbf{u}f_{p}\left(\sqrt{2}\lambda\pi \lvert \textbf{u}\rvert \right) \exp\left(-2i\pi\textbf{u}.\textbf{r}\right)+\frac{g_{p}\left(r/\sqrt{2}\lambda\right)}{r^{p}},
\label{rinv}  
\end{equation} 
where
\begin{equation}
g_{p}(x)=\frac{2}{\Gamma\left(p/2\right)}\int_{x}^{\infty } s^{p-1}\exp (-s^{2})ds 
\label{gp}
\end{equation}
\begin{equation}
f_{p}(x)=\frac{2x^{p-3}}{\Gamma\left(p/2\right)}\int_{x}^{\infty } s^{2-p}\exp (-s^{2})ds,
\label{fp} 
\end{equation}
and $ \lambda $ is an arbitrary parameter. The first and second terms of equation (\ref{rinv}) express the “long-range part” and the “short-range part”, respectively.

\subsection{Polyakov Loops}
 Considering the super cell and its copies and applying a periodic boundary condition, the infinite sum $ \Phi $ in equation (\ref{phi})
\begin{equation}
\Phi(\textbf{r})\equiv \sum_{\textbf{n}\in \mathbb{Z} ^{3}}\sum_{i=1}^{n_{D}}\frac{q_{i}}{\lvert \textbf{r}-\textbf{r}_{i}-\textbf{n}L\rvert }, 
\label{ewaldphi}
\end{equation}
where $ \textit{L}^{3} $ is the spatial volume of the super cell. Putting $ p=1 $ in equation (\ref{rinv}) and calculating $ g_{p} $ and $ f_{p} $, $ \Phi $ is split into a “short-range part” $ \Phi^{\texttt{short}} $ and a “long-range part” $ \Phi^{\texttt{long}} $
\begin{equation}
\Phi(\textbf{r}) =\Phi ^{\texttt{short}}(\textbf{r})+\Phi ^{\texttt{long}}(\textbf{r})
\end{equation}
\begin{equation}
\Phi ^{\texttt{short}}(\textbf{r})\equiv \sum_{\textbf{n}\in \mathbb{Z} ^{3}}\sum_{i=1}^{n_{D}}\left( 1-erf\left( \frac{\lvert \textbf{r}-\textbf{r}_{i}-\textbf{n}L\rvert}{\sqrt{2}\lambda}\right) \right) \frac{q_{i}}{\lvert \textbf{r}-\textbf{r}_{i}-\textbf{n}L\rvert }
\label{short}
\end{equation}
\begin{equation}
\Phi ^{\texttt{long}}(\textbf{r})\equiv \sum_{\textbf{n}\in \mathbb{Z} ^{3}}\sum_{i=1}^{n_{D}}erf\left( \frac{\lvert \textbf{r}-\textbf{r}_{i}-\textbf{n}L\rvert}{\sqrt{2}\lambda}\right) \frac{q_{i}}{\lvert \textbf{r}-\textbf{r}_{i}-\textbf{n}L\rvert },
\label{long}
\end{equation}
where \textit{erf} denotes the error function. Since $ \Phi^{\texttt{short}} $ is exponentially decaying, it is converged for a finite cutoff. Although $ \Phi^{\texttt{long}} $ is a divergent quantity, it is a smooth function. Therefore, its Fourier transformed is converged for a finite cutoff 
\begin{align}
\Phi ^{\texttt{long}}(\textbf{r})=\frac{4\pi}{L^{3}}\sum_{\textbf{n}\in \mathbb{Z} ^{3} \setminus  \vec{0}} \frac{e^{-\lambda ^{2}\textbf{k}(\textbf{n})^{2}/2}}{\textbf{k}(\textbf{n})^{2}}Re\left( \sum_{j=1}^{n_{D}} q_{j}e^{+i\textbf{k}(\textbf{n})\textbf{r}}e^{-i\textbf{k}\textbf{}(\textbf{n})\textbf{r}_{j}}\right) , \textbf{k}(\textbf{n})=\frac{2\pi }{L}\textbf{n} .
\label{ploop}
\end{align}
where 
\begin{equation}
S(k)= \sum_{j=1}^{n_{D}} q_{j}e^{-i\textbf{k}\textbf{}(\textbf{n})\textbf{r}_{j}} 
\label{sfactor}
\end{equation}
 is the structure factor. It should be noticed that this expression is correct for the long-range term because of neutrality of the system.

\subsection{Finite Volume Effect Under Control using Ewald's method}
\begin{table}[b]\footnotesize
\captionsetup{font=footnotesize}
 \begin{center}
    \begin{tabular}{| l | l | l | l |}
    \hline
    $ n_{D} $ & $ LT $ & configurations  \\ \hline
    1000 & 10 & 1600  \\ \hline
    8000 & 20 & 800 \\ \hline
    27000 & 30 & 120 \\ \hline
    64000 & 40 & 90 \\ \hline
    125000 & 50 & 60 \\ \hline
    \end{tabular}
\end{center}
    \caption{input data of reference \cite{3}: the number of dyon configurations, dyon numbers, $n_D$, and $LT$ for each simulation. $L^3$ indicates the spatial volume of the super cell and T is the temperature}
      \label{tab:input}
\end{table}
In this subsection, the results of Bruckmann's and \textit{et al}. \cite{3} are represented briefly. Applying Ewald's method, they computed the free energy of a static quark-antiquark pair as a function of their separation in a non-interacting dyon gas. They fixed the dyon density $ \rho $ and the temperature $ T $ such that $ \rho /T^{3} = 1 $. The simulations were done for different volumes and dyon numbers. The number of dyon configurations and dyon numbers $n_{D}$ for each simulation are listed in table \ref{tab:input}. The main motivation, as they claimed, was to systematically control the finite volume effects in observables such as the Polyakov loop. 

\begin{figure}
 \captionsetup{font=footnotesize}
  \begin{center}
    \includegraphics[width=0.6\linewidth]{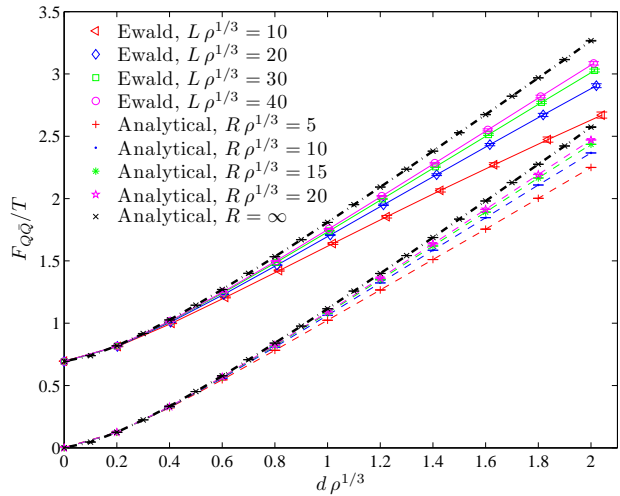}
    \caption{Free energy of a static quark-antiquark pair as a function of their separation for $\rho /T^{3}=1.0$ and various super cell extensions $L\rho ^{1/3}$ corresponding to different dyon numbers $n_{D}$. They showed the results obtained from a numerical evaluation of the analytic result at finite and infinite volume, as well. The analytic results were shifted by $\log2$ and therefore the corresponding curves start close to the origin \cite{3}.}
       \label{fig:bruckmann}
  \end{center}
\end{figure}

The results for both analytical and numerical calculations are illustrated in figure \ref{fig:bruckmann}. Their results are parameterized by
\begin{equation}
 \frac{\sigma }{T^{2}} = \frac{\sigma(T=0)}{T_{c}^{2}}\left(\frac{T_{c}}{T}\right)^{2}A\left(1-\frac{T}{T_{c}}\right)^{0.63}\left(1+B\left(1-\frac{T}{T_{c}}\right)^{1/2}\right),
 \label{scale}
 \end{equation} 
where $B = 1 - 1/A$ and $A=1.39$ \cite{3}. Using lattice result, $\sigma (T=0)=\left(440 \textsl{MeV}\right)^{2}$ which corresponds to $T_{c}=312 $ MeV, they showed the free energy of a quark-antiquark pair increases linearly by increasing the distances between the quark and antiquark. When dyon number $n_{D}$ or $LT$ increases, $\sigma /T^{2}$ which shows the slope of the linear part, converges to $\pi /2$. This is expected as approved by analytical calculation \cite{3}.
\[ \frac{\sigma }{T^{2}}=\frac{\pi }{2}\frac{\rho }{T^{3}}, \]
where $ \rho /T^{3} = 1 $ was fixed in the simulations. As shown in the figure, using Ewald's method, finite volume effect is under control and by choosing large enough super cell , the Ewald's method and analytical and extrapolation to infinite volume results nicely agree within the errors.

In the next section, we explain the particle mesh Ewald \cite{9} which is another numerical method to calculate the free energy between a quark and antiquark . 

\section{ Particle Mesh Ewald's method}
\label{sec:pme}

Although the basic idea of particle mesh Ewald \cite{9} -PME- is as the same as the simple Ewald's method, there is one important difference in assigning the charges to the mesh. This idea was introduced by Hockney and Eastwood \cite{10} in a computer simulation method. In this method, the super cell is gridded after all particles are distributed, and the charge of each particle is assigned to the nearest grid points. In the following, we briefly summarize PME method based on the paper, “A smooth particle mesh Ewald method” \cite{9}. In this paper, only the “long-range part” of the action was simulated by SPME and the “short-range part” was done by simple Ewald's method. Both piecewise Lagrangian and cardinal B-spline interpolations were described in \cite{9}. However, the cardinal B-spline interpolation was applied to compute the reciprocal energy of the system. They used this interpolation since the coefficients $M_{n}(u)$ were $n-2$ differentiable analytically if one needed to compute the reciprocal force and stress tensors. This is unlike the coefficients of piecewise Lagrangian interpolation which are only piecewise differentiable.
\\We follow piecewise Lagrangian interpolation to assign the charges of dyons to the mesh, since we do not need to compute the force and the differentiability. After assigning the new charges on the grid points, we do our simulations by the method introduced in section \ref{sec:em}.

Consider $ n_{D} $ dyons at positions $ \textbf{r}_{1},\textbf{r}_{2},...,\textbf{r}_{n_{D}} $ in the super cell. The vectors $ \textbf{a}_{\alpha } $ ,$ \alpha =1,2,3 $ form the super cell and the conjugate reciprocal vectors $ \textbf{a}_{\alpha }^{*} $ are defined by the relations $\textbf{a}_{\alpha }^{*}.\textbf{a}_{\beta }=\delta _{\alpha \beta} $, $ \alpha,\beta =1,2,3 $. The dyon at position $ \textbf{r}_{i} $ in real space has the fractional coordinates $ s_{\alpha i} $ in reciprocal space, where $ s_{\alpha i}=\textbf{a}_{\alpha }^{*}.\textbf{r}_{i} $. Therefore one can rewrite the structure factor in equations (\ref{sfactor}) in new fractional coordinates, since $ \textbf{m} $'s are the reciprocal lattice vectors, $ \textbf{m}=m_{1}\textbf{a}_{1}^{*}+m_{2}\textbf{a}_{2}^{*}+m_{3}\textbf{a}_{3}^{*} $,
\begin{equation}
S(\textbf{m})=\sum_{i=1}^{n_D}q_{i}\exp \left( -i\textbf{m}.\textbf{r}_{i}\right)=\sum_{i=1}^{n_D}q_{i}\exp \left[-i\left(m_{1}s_{1i}+m_{2}s_{2i}+m_{3}s_{3i} \right)  \right].
\label{structuref} 
\end{equation}
Now, the reciprocal space is gridded to $ K_{1},K_{2},K_{3} $ in all three directions, and the fractional coordinates are scaled to new coordinates $ u_{\alpha }=K_{\alpha }\textbf{a}_{\alpha }^{*}.\textbf{r} $, $ \alpha=1,2,3 $, where $ 0\leq u_{\alpha } < K_{\alpha } $, because of the periodic boundary conditions. Then
\begin{equation}
\exp \left( -i\textbf{m}.\textbf{r}_{i}\right)=\exp \left(-i\frac{m_{1}u_{1}}{K_{1}} \right).\exp \left(-i\frac{m_{2}u_{2}}{K_{2}} \right).\exp \left(-i\frac{m_{3}u_{3}}{K_{3}} \right).
\end{equation}
Using piecewise Lagrangian interpolation, one can approximate these exponential, for $p> 1$
\begin{equation}
\exp \left(-i\frac{m_{\alpha}}{K_{\alpha}}u_{\alpha}\right) \approx \sum _{k=-\infty}^{\infty}W_{2p}(u_{\alpha}-k). \exp\left(-i\frac{m_{\alpha}}{K_{\alpha}}k\right)
\label{exp}
\end{equation}
where $W_{2p}(u^{'})=0$ for $|u^{'}| > p$ and for $-p\leq u^{'} \leq p$ the coefficient $W_{2p}(u^{'})$ is
\begin{equation}
W_{2p}(u^{'}) = \frac{\prod _{j=-p,j\neq k^{'}}^{p-1} (u^{'}+j-k^{'})}{\prod _{j=-p,j\neq k^{'}}^{p-1} (j-k^{'})}, k^{'}\leq u^{'}\leq k^{'}+1, k^{'}=-p,-p+1,...,p-1.
\label{W2p}
\end{equation}
The order of interpolation, 2p, is the number of mesh points used to interpolate $\exp(-i mu/K)$. These points are $[u]-p+1$, $[u]-p+2$ , ..., $[u]+p$, the nearest mesh points to the point u. Now, the structure factor in equation (\ref{structuref}) can be approximated, using equation (\ref{exp}),
\begin{equation}
\begin{split}
S(m)\approx & \widetilde{S}(m)=\sum _{i=1}^{n_{D}}q_{i} \sum _{k_{1}=-\infty}^{\infty}\sum _{k_{2}=-\infty}^{\infty}\sum _{k_{3}=-\infty}^{\infty} W_{2p}(u_{1i}-k_{1})W_{2p}(u_{2i}-k_{2}) {}\\
&.W_{2p}(u_{3i}-k_{3})\exp \left(-i\frac{m_{1}}{K_{1}}k_{1}\right)\exp \left(-i\frac{m_{2}}{K_{2}}k_{2}\right)\exp \left(-i\frac{m_{3}}{K_{3}}k_{3}\right). 
\end{split}
\label{stilda}
\end{equation}
Comparing equations (\ref{structuref}) and (\ref{stilda}), one can find the new charges assigned on mesh
\begin{equation}
   \begin{split} 
Q(k_{1},k_{2},k_{3})=  \sum_{i=1}^{n_D}\sum_{n_{1},n_{2}n_{3}}q_{i} & W_{2p}(u_{1i}-k_{1}-n_{1}K_{1})W_{2p}(u_{2i}-k_{2}-n_{2}K_{2}) {}\\
& .W_{2p}(u_{3i}-k_{3}-n_{3}K_{3}).
   \end{split} 
   \label{Q} 
\end{equation}
Using equation (\ref{Q}) in equation (\ref{stilda}), the structure factor is
\begin{equation}
S(\textbf{m})\approx \sum_{k_{1}=0 }^{K_{1}-1 }\sum_{k_{2}=0 }^{K_{2}-1 }\sum_{k_{3}=0 }^{K_{3}-1 } Q(k_{1},k_{2},k_{3}) \exp \left[-i\left(\frac{m_{1}k_{1}}{K_{1}}+\frac{m_{2}k_{2}}{K_{2}}+\frac{m_{3}k_{3}}{K_{3}} \right) \right].
\label{newstructurefac} 
\end{equation}
The structure factor in equation (\ref{newstructurefac}) describes the system of $K_{1}K_{2}K_{3}$ charges - the number of mesh points in 3D lattice - located on mesh points, $k_{i}$, which is the approximation of the system of $n_{D}$ charges located random in $r_{i}$. Therefore we can work with this new system and use the simple Ewald simulation to compute both short-range and long-range parts of energy and Polyakov loops. The advantage of this new system is that for each number of dyons, $ n_{D} $, we have the specific and constant number of charges, $K_{1}K_{2}K_{3}$.

\section{Numerical Results}
\label{sec:results}
We calculate the free energy of a static quark-antiquark pair as a function of their separation using the Polyakov loop correlator of section \ref{sec:dyon}. We use the Ewald's method introduced in section \ref{sec:em} with charges obtained by PME method obtained in section \ref{sec:pme}. To plot the energy of the static quark-antiquark pair versus distance, we follow the same procedure as reference \cite{3} and the setup of table \ref{tab:input} by fixing the dyon density $ \rho $ and temperature $ T $ to $ \rho /T^{3} = 1 $ which scales the separations by $ \rho ^{1/3} $ or $T$. The number of mesh points are fixed on super cells of different volume in our calculations.

\begin{figure}[t]
\captionsetup{font=footnotesize}
\centering
\begin{subfigure}{.45\textwidth}
  \centering
  \includegraphics[width=1\linewidth,height=1\linewidth]{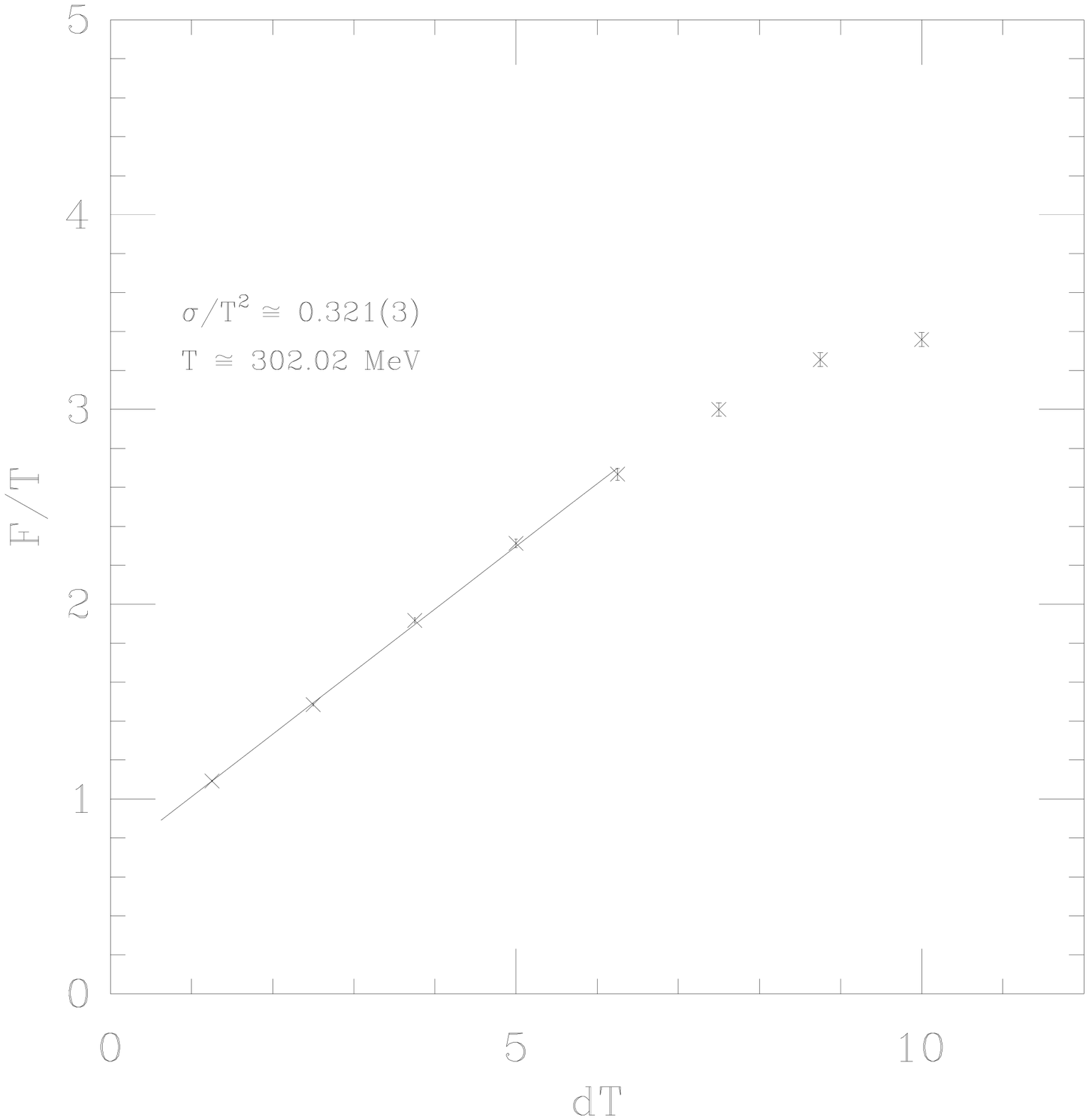}
  \vspace{-40pt}
  \caption{LT = 20}
  \label{fig:20}
\end{subfigure}%
\hspace{30pt}
\begin{subfigure}{.45\textwidth}
  \centering
  \includegraphics[width=1\linewidth,height=1\linewidth]{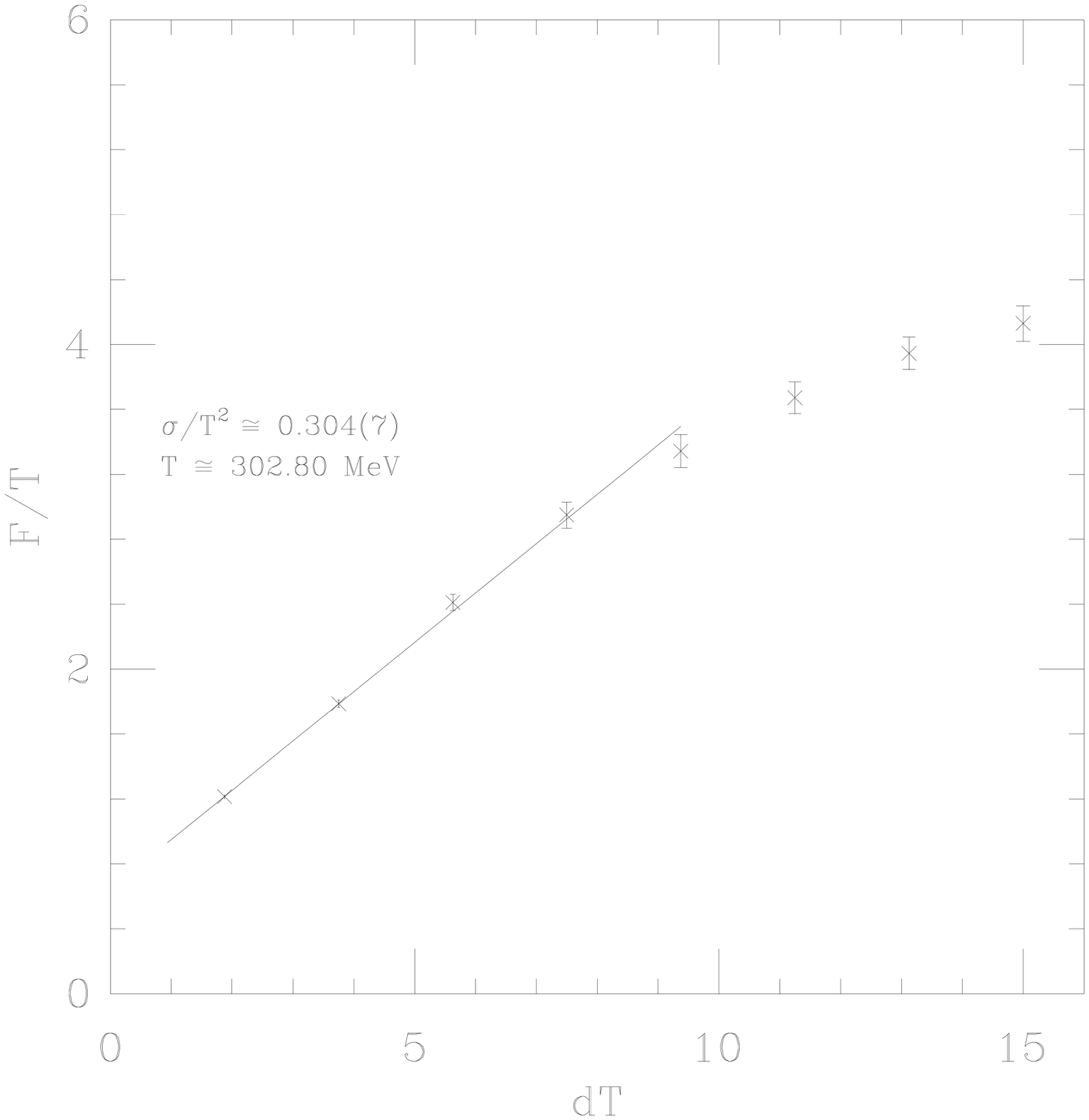}
  \vspace{-40pt}
  \caption{LT = 30}
  \label{fig:30}
\end{subfigure} %\vspace{-40pt}
\caption{The linear dependence of free energy to the quark-antiquark separation for $\rho /T^{3}=1$. The temperatures for our simulations are near the deconfinement temperature, $T=312 $ MeV. The points for larger $dT$ are most probably affected by finite size volume effect and are not used for the linear fit.}
\label{fig:2030}
\end{figure}

As mentioned in section \ref{sec:dyon} we do our calculations for maximally non-trivial holonomy corresponding to confinement phase. Thus, we expect the potential to grow linearly by increasing the quark-antiquark distances.  Figure \ref{fig:2030} illustrates the linear dependence of free energy to the quark-antiquark separation for $LT = 20$ and $LT = 30$  for small and intermediate distances. 
\begin{table}
\captionsetup{font=footnotesize}
{\footnotesize
\begin{tabular}{l|l*{6}{c}r}
 LT & $\sigma /T^{2}$ & T (MeV)& $\sigma (fm^{-2}) $ & lattice spacing (fm) & $\sigma (T)/\sigma (T=0)$ & $T/T_{c}$\\
\hline
 10 & 0.494(5) & 293.28 & 1.1(1) & 0.42 & 0.22 & 0.940   \\ 
 20 & 0.321(3) & 302.02 & 0.76(1) & 0.81 & 0.15 & 0.968   \\ 
 30 & 0.304(7) & 302.80 & 0.72(1) & 1.21 & 0.14 & 0.970   \\ 
 40 & 0.25(1) & 305.14 & 0.60(1) & 1.61 & 0.12 & 0.978    \\ 
 50 & 0.28(1) & 303.89  & 0.67(1) & 2.02 & 0.13 & 0.974   \\
\end{tabular}
}
\caption{The numerical results of our simulations for different LT. The temperatures of these simulations are near critical temperature $T=312 $ MeV and the values of $\sigma (T)/\sigma (T=0)$ are consistent with $\sigma (T)/\sigma (T=0) \approx 0.13$ for $T/T_{c}=0.98$ in paper \cite{11}. The string tensions of the same temperature obtained from the lattices with different lattice spacings agree within the errors. This shows that the discretization error does not affect our results.}
\label{tab:alldata}
\end{table}

\begin{figure}
\captionsetup{font=footnotesize}
  \begin{center}
    \includegraphics[width=.6\linewidth]{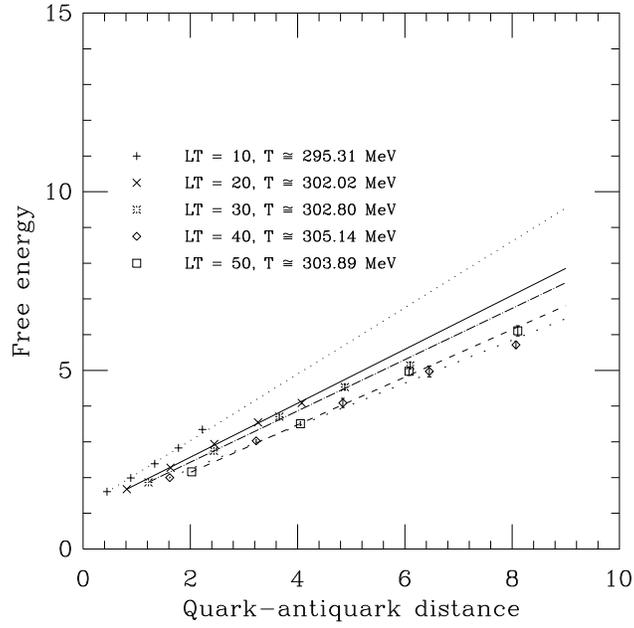}
    %\vspace{-45pt}
    \caption{The scaled results for different volume. As the temperature increases, string tension decreases. For approximately identical temperatures string tensions are close enough to ignore the effect of discreteness.}
       \label{fig:sall}
  \end{center}
  \end{figure}

To compare the different volume simulation results, we should find the temperature to scale our data. We parametrize our results by equation (\ref{scale}). First, the slope $\sigma /T^{2}$ from the fitting of the plot of free energy versus distances is obtained and then using equation (\ref{scale}) the temperature $T$ is obtained on the right hand side. As indicated in figure \ref{fig:sall}, temperatures are $293.28$, $302.02$, $302.80$, $305.14$, and $303,89$ MeV for $LT=10, 20, 30, 40,$ and $50$, receptively. The temperatures are very close to each other. Compared with the critical temperature $T_{c}=312$ MeV, we are very close to the deconfined phase. Given the temperatures, we can find the spatial lattice spacings for each lattice which are $0.42$, $ 0.81 $, $ 1.21 $, $ 1.61 $, $ 2.02 $ fm, for $LT=10, 20, 30, 40,$ and $50$, receptively. In general, in our simulations we get closer to the continuum for smaller $LT$. We would like to recall that since we have used a model 
 where we put charges on a lattice, we have to show that our lattice spacing does not affect the results. In other words, we should show that we are using a lattice spacing which is small enough to not encounter the discretization error. Therefore, the string tensions of the same temperature obtained from the lattices with different lattice spacings should be equal. Table \ref{tab:alldata} confirms that for $LT=20$ and $30$ for which the temperatures are almost equal and also for $LT=40$ and $50$, the string tensions agree within the errors.

\begin{figure}
\captionsetup{font=footnotesize}
  \begin{center}
    \includegraphics[width=0.6\linewidth]{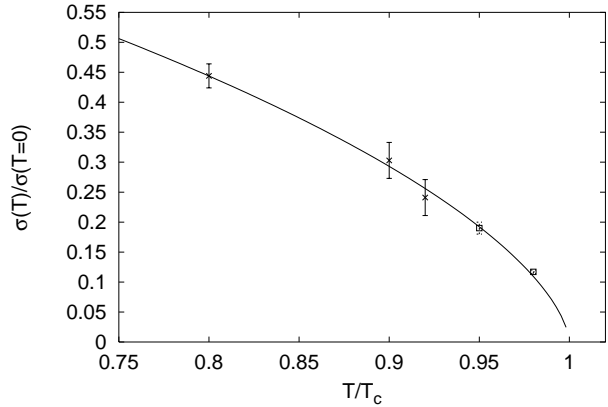}
    \caption{String tension as a function of $T/T_{c}$ \cite{11}. As the temperature approaches the critical temperature, string tension decreases and goes to zero. Our simulations agree this plot and the string tensions decrease by increasing the temperature as indicated in figure \ref{fig:sall}. }
       \label{fig:sigma}
  \end{center}
\end{figure}

On the other hand, string tension changes by temperature as expected from equation (\ref{scale}) and supported in literature like reference \cite{11}. Figure \ref{fig:sigma} shows decreasing the slope of the best fits or $\sigma (T)/\sigma (T=0)$ as the temperature increases or $T/T_{c}$ approaches to one. Our results in figure \ref{fig:sall} indicates the slope of the best fits decrease as the temperature increases. The details of our results given in table \ref{tab:alldata} show the good consistency with $\sigma (T)/\sigma (T=0) \approx 0.13$ for $T/T_{c}=0.98$ in paper \cite{11}.

\begin{figure}
 \captionsetup{font=footnotesize}
  \begin{center}
    \includegraphics[width=.5\linewidth]{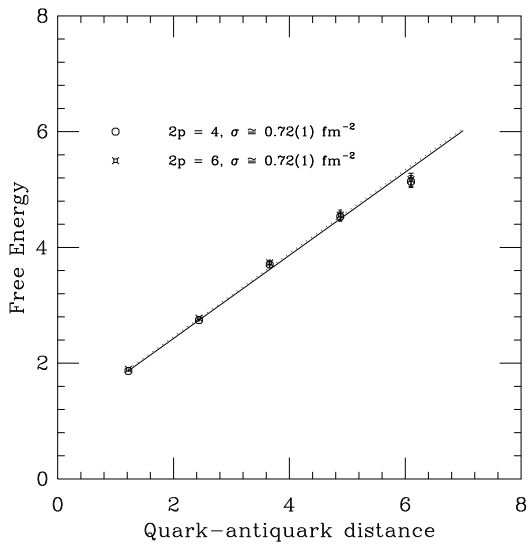}
  %  \vspace{-45pt}
    \caption{The results of $LT = 30$ for order of interpolation $2p = 4$, and $6$. The order of interpolation specifies that the charge of each dyon is interpolated to 2\textit{p} nearest neighbor mesh points in each direction. }
       \label{fig:468}
  \end{center}
\end{figure}

For all these diagrams the order of interpolation, $2p$, in (\ref{exp}) in section \ref{sec:pme}, is fixed to 4. It means that the charge of each dyon is interpolated to four nearest neighbor mesh points in each direction. To examine how proper this choice is, we do the calculations for $LT = 30$ and different order of interpolation as illustrated in figure \ref{fig:468}. It is clear that our choice of $ 2p = 4$ is good enough since the diagrams for different $2p$ tend approximately identical.


\section{Conclusion}
In this work, using the Polyakov correlator we have calculated the free energy of a quark-antiquark pair as a function of their separation by Ewald's method and  Particle Mesh Ewald (PME) method for non-interacting dyon gas. Our results show a linear rising for the potential between quarks for small distances for all temperatures.  In addition, in agreement with lattice results, the string tension decreases by increasing the temperature. 
Using different lattice spacings, we show that discretization does not affect our results. Since we did the calculation for the temperature close to the deconfinement temperature, a qualitative comparison with Ewald's method which was done by Bruckmann and \textit{et al}. \cite{3} is possible. The general behavior of the free energy agrees very nicely with their results even though because of some technical problems we are not able to reach to the temperature they got by the Particle Mesh Ewald's method. In fact we have to enlarge the number of mesh points which is not easily doable with our current computer facilities. 

Particle mesh Ewald's method is less time consuming compared with Ewald's method and therefore must be more efficient in calculating the free energy of a quark-antiquark pair in an interacting ensemble of dyons.

\section{Acknowledgement}

We would like to express our great gratitude to  Michael Muller-Preussker for the very helpful discussions at the XIth Quark Confinement and Hadron Spectrum Conference in Russia. We are very sorry that the Physics society missed such a nice and valuable scientist. We would also like to thank his ex-students Benjamin Maier and Marc Wagner for helping us to understand the parameters of their codes and answering our questions.
 
We are grateful to the research council of the University of Tehran for
supporting this study.

\end{document}